\documentclass[11pt,conference,letterpaper]{IEEEtran}
\usepackage{amsmath}
\usepackage{amstext}
\usepackage{amssymb}
\usepackage[amssymb]{SIunits}
\usepackage{latexsym}
\usepackage{theorem}
\usepackage{placeins}
\usepackage{epsfig}
\usepackage{epstopdf}
\usepackage{xspace}
\usepackage{moreverb}
\usepackage{times}
\usepackage{mathptm}
\usepackage{fancyhdr}
\usepackage{lastpage}
\usepackage{ifthen}
\usepackage{changebar}

\newcommand{\yyear}{{\ensuremath {\mathrm y}\xspace}}
\newcommand{\eV}{{\ensuremath \mathrm{eV}\xspace}}

\newcommand\bra[1]{\langle#1|}
\newcommand\ket[1]{|#1\rangle}
\newcommand\iprod[2]{\langle #1 | #2 \rangle}

\newcommand{\LI}{\begin{itemize}}
\newcommand{\IE}{\end{itemize}}
\newcommand{\LN}{\begin{enumerate}}
\newcommand{\NE}{\end{enumerate}}
\newcommand{\LD}{\begin{description}}
\newcommand{\DE}{\end{description}}

\newcommand{\ie}{i.e.{}\xspace}			
\newcommand{\eg}{e.g.{}\xspace}			
\newcommand{\viz}{viz{}\xspace}			
\newcommand{\cf}{cf.{}\xspace}			
\newcommand{\etc}{etc.{}\xspace}		

\newcommand{\ibid}{\emph{ibid.}}		

\newcommand{\Qed}{$\bold{\Box}$}

\newcounter{mytopic}
\begin{document}
\newboolean{extended}
\setboolean{extended}{false}

\ifthenelse{\boolean{extended}}{
	\lhead{\it Extended document}
}{}
\nochangebars

\title{\Large\sc
	A wave effect enabling
	universal frequency scaling,
	monostatic passive radar,
	incoherent aperture synthesis,
	and
	general immunity to jamming and interference
	}
\author{
		V. Guruprasad\\
		Inspired Research, New York.
	}
\maketitle
\pagestyle{fancy}
\thispagestyle{fancy}
\noindent
%
\section*{ABSTRACT}

\noindent
\textit{
A fundamental Doppler-like but asymmetric wave effect
that shifts
	received signals in frequency
in proportion to
	their respective source distances,
was recently described as means for
	a whole new generation of communication technology
using angle and distance,
	potentially replacing
	TDM, FDM or CDMA,
		for multiplexing.
It is equivalent to
	wave packet compression
by scaling of time
	at the receiver,
converting
	path-dependent phase into
		distance-dependent shifts,
and can multiply
	the capacity of physical channels.
The effect was
	hitherto unsuspected in physics,
appears to be responsible for both
	the cosmological acceleration
and
	the Pioneer 10/11 anomaly,
and is exhibited
	in audio data.
This paper discusses
	how it may be exploited for
instant, passive ranging of signal sources, for
	verification, rescue and navigation;
incoherent aperture synthesis
	for smaller, yet more accurate radars;
universal immunity to
	jamming or interference;
and
	precision frequency scaling of
		radiant energy in general.
}


\newcommand{\Section}[2]{\section{#2}\label{s:#1}}
\newcommand{\Subsection}[2]{\subsection{#2}\label{ss:#1}}
\newcommand{\unc}{}
\newcommand{\snf}{}
\newcommand{\con}{}
\newcommand{\func}{}
\newcommand{\fsnf}{}
\newcommand{\fcon}{}
\Section{intro}{Introduction} 

\unc
A previously unsuspected result of the wave equation,
enabling a receiver to get frequency shifts
	in an incoming signal
in proportion to
	the physical distance of its source,
		was recently described
	\cite{Prasad2005}.
Its main premise,
that signals from real sources
	must have nonzero spectral spread,
seems to be well supported by
	the cosmological and
	the Pioneer 10/11 anomalous acceleration datasets,
and
	a remarkably large gamut of terrestrial mysteries
		that can be mundanely resolved.
This perceived support actually concerns
a further inference that the mechanism
	could occur naturally in our instruments
on the order of magnitude of
	$
	10^{-18}
	~\reciprocal{\second}
	$.
This is a decay corresponding, as half-life, to
	the age of the solar system,
and is too small for
	purely terrestrial applications.
The wave effect is consistently demonstrated with
	acoustic samples, however,
testifying to
	its fundamental and generic nature.
While a general exposition deserves to be made in due course
	in a physics forum,
it presents fundamental new opportunities for
	intelligence and military technologies.

\unc
Section \ref{s:theory} contains a brief review of
	the theory of the effect,
showing that
	all real signals necessarily carry source distance information
		in the spectral distribution of phase,
analogous to
	the ordinary spatial curvature of wavefronts,
and that by scanning this phase spectrum,
	each frequency in a received waveform
			comprising multiple signals
	would be shifted in proportion to
		its own source distance and the scanning rate.
Section \ref{s:impl} presents
	the general principles of realization
by both
	spectrometry
and
	digital means.
Section \ref{s:formalism} shows
	how this enables separation of signals by physics
		instead of modulation,
whose implications for information theory
	have been discussed [\ibid].
Intelligence and military possibilities
	are considered for the first time
		in Section \ref{s:mil}.


\Section{theory}{Hubble's law shifts, temporal parallax} 

\begin{figure}[ht]
	\centering
	\psfig{file=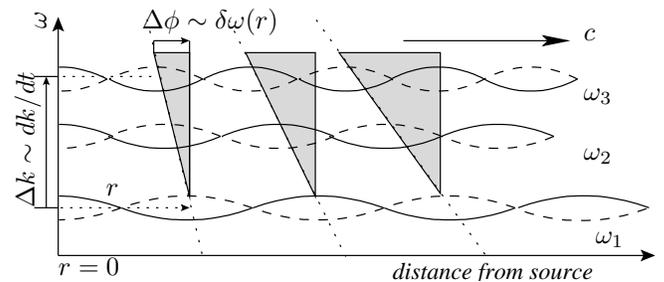}
	\caption{Phase contours and gradient for a source impulse}
	\label{f:PhaseGradient}
\end{figure}

\unc
The sole premise for the effect, as mentioned, is
	the nonzero bandwidth of a real signal.
The Green's function for the general wave equation
	concerns an impulse function
		$\delta(\mathbf{x},t)$
	as the elemental source.
Its Fourier transform is
\begin{equation} \label{e:impulseFT}
	F[\delta(t)] = 
		\int \delta(t) \, e^{-i \omega t} \, dt
	=
		1
	\quad
	,
\end{equation}
which says that
\emph{all spectral components of an impulse
	start with the same phase}.
The source itself is the only common reference
	across any continuous set of frequencies,
hence the \emph{spectral phase contours}
	indicate its distance
	(Fig.~\ref{f:PhaseGradient}).

\unc
The main difficulty in tapping this available source
	of source distance information 
is that it lies in
	differences of phase across incoming frequencies,
and
	phases are hard to measure accurately.
However,
if we scanned the spectrum at rate $d \widehat{k}/dt$,
then,
	for each contour $\phi$
	and source distance $r$,
we should encounter
	an increasing or decreasing phase
proportional to
	$r$ and
	the integration interval $\Delta \widehat{k}$,
as shown by
	the shaded areas in the figure.
The trick that yields
	this remarkable asymmetric effect
is in the definition that
	\emph{a rate of change of phase is a frequency},
so that
	the instantaneous measures of the spectral phase contours
		obtained from the scanning
have the form of
	frequency shifts proportional to the slopes,
and therefore to
	the respective source distances.
These distance-dependent shifts
	resemble Hubble's law in astronomy
but are strictly linear
	due to their nonrelativistic, mundane origin,
and are rigorously predicted
	by basic wave theory,
		as follows.
The instantaneous phase of a sinusoidal wave
	at $(\bold{r}, t)$,
from a source at $\mathbf{r} = 0$,
	would be
\begin{equation} \label{e:twPhase}
	\phi(\mathbf{k}, \omega, t) =
		\mathbf{k} \cdot \mathbf{r} - \omega \; t
	\quad
	,
\end{equation}
and leads to
	the differential relation
\begin{equation} \label{e:dphase}
	\left. \Delta \phi \right|_{\omega,t}
	=
		\Delta (\omega \; t)
	+
		\mathbf{k} \cdot \Delta \mathbf{r}
	+
		\Delta \mathbf{k} \cdot \mathbf{r}
	\quad
	.
\end{equation}
The first term
	$\Delta (\omega \; t)$
clearly concerns the signal content if any,
	and has no immediate relevance.
The second term
	$\mathbf{k} \cdot \Delta \mathbf{r}$
expresses
	the path phase differences
		at any individual frequency,
and is involved in
	the ordinary Doppler effect,
as it describes phase change
	due to changing source distance,
as well as
	image reconstruction in holography
		and synthetic aperture radar (SAR),
as the reconstructed image concerns
	information of incremental distances of
		the spatial features of the image.
The third term,
	clearly orthogonal to other two,
represents
	phase differences across frequencies
		for any given (fixed) source distance.
Then, by varying a frequency selection $\widehat{k}$
	at the receiver,
we should see
	an incremental frequency, or shift,
\begin{equation} \label{e:fshift}
	\delta \omega =
		\lim_{\Delta t \rightarrow 0}
		\left[
			\frac{
				\left.
				\Delta \phi(\omega)
				\right|_{\omega,r,t}
			}
			{ \Delta k }
		\times
			\frac{ \Delta \widehat{k} }
			{ \Delta t }
		\right]
	=
		\left.
			\frac{\partial \phi}{\partial k}
		\right|_{\omega,r,t}
		.
			\frac{d \widehat{k}}{dt}
	\quad
	.
\end{equation}
(Note that $k$ can only refer to the incoming wave vector
	in the denominator.)
As this holds for each individual $\omega$,
	from equation (\ref{e:twPhase}),
the signal spectrum would be uniformly shifted by
	the normalized shift factor
\begin{equation} \label{e:zshift}
\begin{split}
	z(r) \equiv
		\frac{\delta \omega}{\widehat{\omega}}
	&=
		\frac{1}{\widehat{\omega}}
		\left.
			\frac{\partial \phi}{\partial k}
		\right|_{\omega,r,t}
		.
			\frac{d \widehat{k}}{dt}
	=
		\frac{r}{\widehat{k} c}
		.
			\frac{d \widehat{k}}{dt}
	=
		\frac{\beta r}{c}
	=
		\alpha \, r
	\quad
	,
\\
\text{with}\quad
	\beta &\equiv \widehat{k}^{-1} (d \widehat{k}/dt)
\quad
\text{and}
\quad
	\alpha = \beta / c
	\quad
	,
\end{split}
\end{equation}
	where $t$ denotes time kept
		by the receiver's clock.
Further,
the shifts reflect only
	the instantaneous value of $\beta$,
which is the normalized rate of change
		at the receiver.

\begin{figure}[ht]
	\centering
	\psfig{file=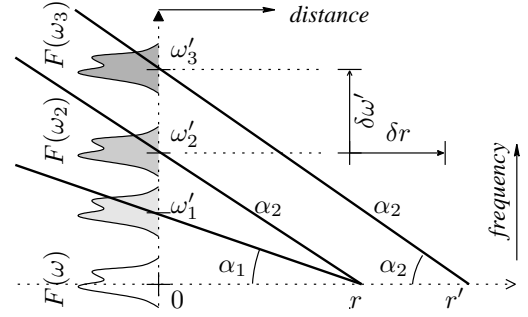}
	\caption{Temporal parallax}
	\label{f:Parallax}
\end{figure}

\unc
Fig.~\ref{f:Parallax} shows
that $\delta \omega$ is
	a temporal analogue of spatial parallax,
with the instantaneous value of $\alpha \equiv \beta/c$
serving in the role of
	lateral displacement of the observer's eyes:
at each value
	$\alpha_1$ or $\alpha_2$,
the signal spectrum $\mathcal{F}(\omega)$ shifts to
	$\mathcal{F}(\omega_1)$ or
	$\mathcal{F}(\omega_2)$,
	respectively,
and if the distance to the source were increased to
	$r + \delta r$,
the spectrum would further shift to
	$\mathcal{F}(\omega_3)$.
Unlike with ordinary (spatial) parallax,
	the receiver can be fixed \emph{and monostatic},
and yet exploit
	the physical information
		of source distance.

\Section{impl}{Physics of realization} 

\unc
There are three basic ways of accomplishing
	Fourier decomposition or frequency selection
in a receiver: using
		diffraction or refraction,
		a resonant circuit,
	or 
		by sampling and digital signal processing (DSP).
Realization of the shifts in all three methods
	has been described,
along with how this ``virtual Hubble'' effect
	had remained unnoticed so long
	[\ibid].
A review of 
	the diffraction and DSP methods
is necessary as theoretical foundation for
	the application ideas to be discussed.
These methods would also be applicable for
	array antennas and software-defined radio,
		respectively.

\Subsection{diffract}{Diffractive implementation} 

\begin{figure}[ht]
	\centering
	\psfig{file=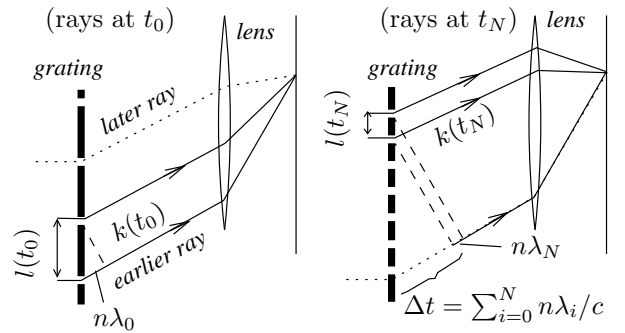}
	\caption{Time-varying diffraction method}
	\label{f:GratingInstants}
\end{figure}

\unc
Diffractive selection of
	a wavelength $\widehat{k}$
concerns using a diffraction grating
	to deflect normally incident rays
to an angle $\theta$
	corresponding to the selected $\widehat{k}$,
as determined by the grating equation
	$n \lambda = l \sin \theta $,
	where $n$ denotes the order of diffraction,
		and $l$ is the grating interval,

\unc
The property exploited is that
rays arriving at one end of the grating
combine with rays
	that arrived at the other end
		\emph{a little earlier}.
If we could change the grating intervals $l$
	inbetween,
the rays that get summed at
	a diffraction angle $\theta$
		would correspond to changing
			grating intervals $l(t)$,
as depicted in
	Fig.~\ref{f:GratingInstants}.
Their wavelengths must then relate to
	the grating interval as
	$
	n \, d \lambda/dt = (dl/dt) \, \sin \theta 
	$.
Upon dividing this by
	the grating equation,
we obtain
	the modified relation
	$
	\widehat \lambda^{-1} (d\widehat \lambda/dt)
	=
	\widehat l^{-1} (d\widehat l/dt)
	\equiv
		- \beta .
	$

\unc
This result is not merely
	a coincidence of different wavelengths
		at the focal point,
as would be obtained
	with a nonuniform grating.
Each of the summed contributions itself
	has a time-varying $\lambda$,
so the variation over the sum is consistent
	with the component variations.
A grating gathers more total power
	than, say, a two-slit device,
and the waves arrive with
	instantaneously varying phases
\begin{equation} \label{e:phaserate}
		\frac{d \phi}{dt}
	\equiv
		- \widehat\omega
	=
		\frac{\partial \phi}{\partial t}
	+
		\nabla_\mathbf{r} \phi \cdot \mathbf{\dot{r}}
	+
		\nabla_\mathbf{r'} \phi \cdot \mathbf{\dot{r}'}
	+
		\nabla_{\mathbf{k}} \phi \cdot
			\mathbf{\dot{\widehat{k} }}
	\quad
	,
\end{equation}
where
	the first term $\partial \phi/\partial t$ is
		the signal ($\omega \; t$) contribution
			in equation (\ref{e:twPhase})
		and is equal to $-\omega$;
	the second term (in $\mathbf{\dot{r}}$) concerns
		the real motion and Doppler effect if any,
			to be ignored hereon;
	the third term (in $\mathbf{\dot{r}'}$) represents
		a similar Doppler shift from
			longitudinal motion of	
				either end of the receiver,
		which would be generally negligible
			for $r' \ll r$;
	and
	only the last term concerns
		equation (\ref{e:fshift}).

\Subsection{dsp}{DSP implementation} 

\unc
In DSP,
$\widehat{k}$'s are determined by
	the sampling interval $T$,
and can therefore be continuously varied
	by changing $T$.
The discrete Fourier transform (DFT)
	is defined as
\begin{equation} \label{e:dft}
\begin{split}
	F(m \omega_T)
	&=
		\sum_{n = 0}^{N - 1}
			e^{i m \omega_T T}
			\,
			f(n T)
\\
\text{with the inverse}
\quad
	f(n T)
	&=
		\frac{1}{N}
		\sum_{m = 0}^{N - 1}
			e^{i m \omega_T T}
			\,
			F(m \omega_T)
	\quad
	,
\end{split}
\end{equation}
where
	$f$ is the input signal;
	$T$ is the sampling interval;
	$N$ is the number of samples per block;
and
	$\omega_T = 2 \pi / N T$.
The inversion is governed by
	the orthogonality condition
\begin{equation} \label{e:dordorth}
	\sum_{n = 0}^{N - 1}
		e^{i m \omega_T T}
		\,
		e^{i (k r - n \omega_T T)}
	=
		\frac{
			1 - e^{i (m - n)}
		}{
			1 - e^{i (m - n)/N}
		}
	=
		N \delta_{mn}
	\quad
	,
\end{equation}
where $\delta_{mn} = 1$
	if $m = n$, else $0$.
These definitions as such suggest that
the instantaneous selections
	$\widehat\omega_T \equiv \widehat k c$
can be varied via
	the sampling interval $T$.

\unc
To verify that
	a controlled variation of $T$
will indeed yield
	the desired shifts,
observe that in equation (\ref{e:phaserate}),
both the real Doppler terms, in
	$\dot{\mathbf{r}}$ 
and
	$\dot{\mathbf{r}}'$,
can be ignored,
	as $r' \ll r$ for sources of practical interest.
The surviving term on the right is
$
	(\partial \phi/\partial k)
	(d\widehat k/dt)
$
where
	$\partial \phi/\partial t = - \omega$,
as before,
and
\begin{equation*}
	\frac{d\widehat k}{dt} =
		\frac{1}{c}
		\frac{
			d \widehat \omega_T
		}{dt}
	=
		\frac{1}{c}
		\frac{d}{dt}
		\left(
			\frac{2 \pi}{NT}
		\right)
	=
		- \frac{2 \pi}{N c T^2}
		\frac{dT}{dt}
	=
		- \widehat k
		\,
		\frac{1}{T}
		\frac{dT}{dt}
	\quad
	,
\end{equation*}
so that, corresponding to equation (\ref{e:zshift}),
	we do get
\begin{equation} \label{e:samplingmod}
	\widehat k^{-1} (d\widehat k/dt) \equiv
		\beta
	=
		- T^{-1} (dT/dt)
	\quad
	,
\end{equation}
confirming the desired effective variation of
	$\widehat{k}$'s.
\Qed

\begin{figure}[h]
	\centering
	\psfig{file=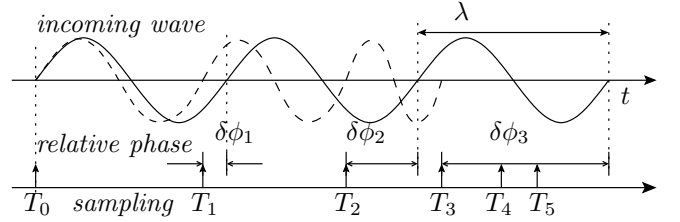}
	\caption{Variable sampling method}
	\label{f:TimeDomain}
\end{figure}

\unc
Fig.~\ref{f:TimeDomain} explains the result.
The incoming wave 
presents increasing phase differences
	$\delta \phi_1$,
	$\delta \phi_2$,
	$\delta \phi_3$, ...
within the successive samples obtained from
	the diminishing intervals
	$\delta T_1 = T_1 - T_0$,
	$\delta T_2 = T_2 - T_1$,
	$\delta T_3 = T_3 - T_2$,
	\etc
From the relation
	$\widehat \omega_T \equiv \widehat k c = 2 \pi / N T$,
gradients of
	the spectral phase contours
		can be quantified as
\begin{equation*}
	\frac{\partial \phi}{\partial T} = 
		\frac{\partial \phi}{\partial \widehat k}
		\frac{d \widehat k}{dT}
	=
		-
		\frac{2 \pi}{N c T^2}
		\frac{\partial \phi}{\partial \widehat k}
\end{equation*}
so that by equation (\ref{e:samplingmod}),
	each phase gradient reduces to
\begin{equation} \label{e:ksampling}
		\frac{\partial \phi}{\partial T}
		\frac{dT}{dt}
	=
		\frac{- 2 \pi}{N c T}
		\frac{\partial \phi}{\partial \widehat k}
	\cdot
		\frac{1}{T}
		\frac{dT}{dt}
	=
		\frac{\widehat \omega_T}{c}
		\frac{\partial \phi}{\partial \widehat k}
	\cdot
		\frac{1}{\widehat k}
		\frac{d\widehat k}{dt}
	=
		\frac{\partial \phi}{\partial \widehat k}
		\frac{d\widehat k}{dt}
\end{equation}
identically,
	validating the approach.

\Subsection{scale}{General operating principles} 

\unc
For a desired shift $z$ at a range $r$,
	equation (\ref{e:zshift}) yields
\begin{equation} \label{e:scaling}
	\frac{\delta \widehat{k}}{k}
\equiv
	- \frac{ \delta {T}}{T}
	\approx
		\frac{ c z \; \delta t}{r}
\end{equation}
for normalized incremental change of
	the receiver's grating or sampling interval
required 
	over a small interval $\delta t$.
In a DSP implementation,
	appropriate at suboptical frequencies,
this determines
	the needed rate of change
		over a sampling interval $\delta T$.
As $\delta T$ would be nominally chosen
based on
	the (carrier) frequencies of interest
		and not on the range,
the nominal rate of change for achieving
	a useful $z$ at a given range $r$
would be independent of
	the operating frequencies.

\unc
For example,
for a $1~\giga\hertz$ signal,
a suitable choice of sampling interval
	is $\delta t = 100~\pico\second$,
		regardless of the range.
Chosing $z = 2$,
which is fairly large in astronomical terms
	but convenient for the present purposes,
we obtain
	$
	\delta \widehat{k}/{k}
		= 6 \times 10^{-2} / r 
	$
per sample
	(taking $c = 3 \times 10^8~\metre\per\second$).
At $r = 1~\metre$,
the result is a somewhat demanding
	$6 \times 10^{-2}$,
	or $6\%$, per sample,
but for $r = 1, ~10$ or $1000 ~\kilo\metre$,
it is
	$
	6 \times 10^{-5},~
	6 \times 10^{-8}
	$
and
	$
	6 \times 10^{-11}
	$
per sample,
	respectively.

\unc
To be noticed is that
since equation (\ref{e:scaling}) prescribes
	a normalized rate of change,
the instantaneous rate of change must grow
	\emph{exponentially}
		in the course of the observation,
since by
	integrating equation (\ref{e:scaling}),
we get
\begin{equation} \label{e:exp}
	\Delta\widehat{k}
\equiv
	\Delta T ^{-1}
	=
		e^{c z \Delta t / r}
	,
\end{equation}
where $\Delta t$ denotes
	the total period of observation.
While an exponential variation is
	generally difficult to achieve
		other than with DSP,
two other problems
	must be also contended with:
the total variation over
	an arbitrarily long observation
		would be impossible anyway
and
it could exceed the source bandwidth even over
	a fairly short observation.

\unc
The only solution is to split
	the observation time into windows,
		as in DFT,
resetting the sampling interval and its variation
	at the beginning of each window,
so that
	the variation is only continuous
		within each window.

\unc
For example,
the normalized rate
	$6 \times 10^{-8}$
calculated above for
	$100~\pico\second$ sampling
and
	$10~\kilo\metre$ range
amounts to
	$3.773 \times 10^{260}$ for
		each second of observation!
Over a $1~\micro\second$ window, however,
	the sampling rate would change
		only by the factor $1.0006$.
This is also
	the spectral spread required of
the wavepackets emitted by the source,
and conversely,
	the window can be set to match
		the spread.

\unc
It is possible to \emph{cascade}
	multiple stages of gratings or DSP,
in order to multiply
	the magnitude of the shifts.
Cascading seems straightforward with
	optical gratings, 
but in DSP,
as the data stream is already broken up
	into discrete samples by the first stage,
and the sampling interval is to be varied
	in each successive stage
		relative to its predecessor,
it becomes necessary to interpolate the samples
	between stages.
Interpolation is similarly needed for
	the reverse shift stage in
		distance-based signal separation
	(Section \ref{ss:ddm}).

\unc
These shifts have been verified only by simulation
for
	electromagnetic waves.
For sound data
	($c \approx 330~\metre\per\second$),
the software does reveal
	nonuniform spreading of
		subbands of acoustic samples,
consistent with
	the likely distributions of their sources.
Testing in an adequately equipped sound laboratory,
	as well as with radio waves,
is still needed.


\Subsection{diff}{Differential techniques} 

\unc
The effect thus scales well
	from terrestrial to planetary distances,
but 
the incremental changes required per sample
	are very small.
Errors in these changes,
	from inadvertent causes
		including thermal or mechanical stresses,
could lead to large errors
	in distance measurements using the effect.
The linearity of the effect can be exploited
to offset such errors by
	differential methods.
From equation (\ref{e:zshift}), we have
\begin{equation} \label{e:zdelta}
	\Delta z
	= 
		\alpha \Delta r
	+
		r \Delta \alpha
	+
		o(\Delta r, \Delta \alpha)
	,
\end{equation}
relating first order uncertainties.
First order error due to
	an uncertainty $\Delta \alpha$
can be eliminated by measurements
	using different $\alpha$
and using the differences.

\unc
Specifically,
an error $\Delta z$ would limit
	the capability for source separation,
		presuming $r$ is known.
By designing a receiver to depend on
	the difference between two sets of shifts
		for the same incoming waves,
\eg by applying the DSP method twice
	with different values of $\alpha$,
a $\Delta z$ error can be eliminated.
Conversely,
	for ranging applications,
where $z$ is the measured variable,
from
	$
	r = z / \alpha
	$,
we get
\begin{equation} \label{e:rdelta}
	\Delta r
	= 
		\Delta z / \alpha
	-
		r \Delta \alpha
	+
		o(\Delta z, \Delta \alpha)
	,
\end{equation}
so that
	the first order error $\Delta r$ can be 
		once again eliminated using differences.
This is 
	a temporal form of triangulation,
illustrated by the lines for
	$\alpha_1$ and $\alpha_2$
that converge at $r$
	in Fig.~\ref{f:Parallax}.
Higher order differences would yield
	more accuracy.


\Subsection{phaseshift}{Phase shifting and path length variation} 

\con
The main difficulty with
	the method of Section \ref{ss:diffract}
is ensuring uniformity of the grating intervals
	even while they are changing.
The method is not realizable by acousto-optic cells
	for this reason.
Controlled nonuniform sampling and sample interpolation 
	pose similar difficulties for DSP%
	\footnote{
	\func
	Radio telescopes, for instance, incorporate
		1 or 3 bit sampling at an intermediate frequency
			for correlation spectroscopy.
	Interpolation would add more noise
		to this already low phase information.
	}.\xspace

\con
The solution is to instead vary the path length
	\emph{after} a fixed grating,
say using the longitudinal Faraday effect
	and circular polarization.
As the variation now concerns a bulk property,
	it would be easier.
The proof is straightforward.
The analogous simplification for DSP is to modify,
	instead of the sampling interval $T$,
the forward transform kernel to
\begin{equation} \label{e:kdft}
	F^{(\omega)} (m \widehat{\omega}_0)
	\equiv
	\sum_{n = 0}^{N - 1}
		e^{i m \widehat{\omega}(t) T}
		\,
		f(nT)
	,
\quad
	\widehat{\omega}(t)
	=
		\widehat{\omega}_0
		\, e^{\beta \, n T}
	,
\end{equation}
\ie
	apply changing phase shifts
		to the successive samples
	corresponding to the path length. 
This is more complex,
but avoids interpolation noise
	and access to the RF frontend.


\Section{formalism}{Source separation theory} 

\Subsection{kernel}{Transformation kernel} 

\unc
Fourier theory more generally involves the continuous form of
the orthogonality condition (\ref{e:dordorth}),
given by
\begin{equation} \label{e:ordorth}
	\int_t
		e^{i \widehat\omega t} \,
		e^{i (k r - \omega t) } \, dt
	=
		e^{ikr} \delta(\widehat\omega - \omega)
	\quad
	,
\end{equation}
where
	$\delta()$ is the Dirac delta function.
All of the methods described require continuously varying
	the mechanism of frequency selection.
This is equivalent, in a basic sense,
	to varying the receiver's notion of
		the scale of time,
the result being
	a change of the orthogonality condition to
\begin{equation} \label{e:modorth}
\begin{split}
	\int_t
		e^{i \widehat\omega \Delta(t)}
		e^{i [kr \Delta(r) - \omega t]}
		\, dt
	&=
	\int_t
		e^{ikr\Delta}
		\,
		e^{i (\widehat\omega \Delta - \omega t)}
		\, dt
\\
	&\equiv
		e^{ikr\Delta}
		\,
		\delta [\widehat\omega \Delta - \omega]
	\quad
	,
\end{split}
\end{equation}
where
	$\Delta \equiv \Delta(r) = (1 + \alpha r)$
	from equation (\ref{e:zshift}),
but is also equivalent to
	$\Delta(t) = (1 + \alpha ct)$,
via the relation 
	$c = r/t$.

\unc
How did we get to equation (\ref{e:modorth})?
Notice that
	equation (\ref{e:ordorth})
already contains
	the frequency selection factor 
		$e^{i \widehat{\omega} t}$
--
the $\Delta$ in the exponent is
	the variation of 
		$\widehat{\omega}$
	provided by the methods of Section \ref{s:impl}.
The second factor
	$e^{i [kr\Delta(r) - \omega t]}$
must correspondingly represent
	the incoming component picked by
		the modified selector;
its only difference from
	equation (\ref{e:ordorth})
is the $\Delta$ multiplying
	the path contribution to phase, $kr$.
Why should this $\Delta$ multiply $kr$
	and not $\omega t$?
The answer lies in the basic premise of
	equations (\ref{e:twPhase})-(\ref{e:zshift})
that our receiver manipulates
	the phase of each incoming Fourier component
		individually,
hence $\omega t$, representing	
	the signal component of the instantaneous phase,
		is not touched.

\unc
The $kr$ term is affected, however,
since the summing builds up
	amplitude at only
		that value of $k$ which matches
	$
	\widehat{\omega} \equiv \widehat{k} c
	$.
This aspect is readily verified by simulation,
and it defines a lower bound on
	the window size (Section \ref{ss:scale})
		in the optical methods,
since the window reset would break
	the photon integration in a photodetector.

\unc
Equation (\ref{e:modorth}) is more general than
	equation (\ref{e:ordorth}),
as the latter corresponds to 
	the special case of $\Delta = 1$
		or $\alpha = 0$.
This property has
	a fundamental consequence
impacting
	all of physics and engineering:
\emph{
Since $\alpha$ refers to
	the physics of the instruments
and could be
	arbitrary small but nonzero,
there is fundamentally no way to rule out 
	a nonzero $\alpha$ in
		any finite local set of instruments.
The only way to detect a nonzero $\alpha$
	is to look for spectral shifts
		of very distant objects
}
as a large enough $r$ would yield
	a measurable $z(r)$%
	\footnote{
	\label{foot:cosmos}
	\func
	A personal hunch that something like this
		could be responsible for
			the cosmological redshifts,
	had prompted informal prediction of
		the cosmological acceleration to
			some IBM colleagues in ca.1995-1996.

	\unc
	The modified eigenfunctions
		$e^{i [kr\Delta(r) - \omega t]}$
	are known in astrophysics
		as the photon and particle eigenfunctions
			over relativistic space-time
		\cite{Parker1969}.
	The natural value of
		[$\alpha \approx$] $10^{-18}~\reciprocal{\second}$
	cited in the Introduction can come from
	the probability factor
		$e^{-W/k_B T}$
	for cumulative lattice dislocations
		under centrifugal or tidal stresses
			causing creep.
	At $T \approx 300~\kelvin$,
	it is
		$3 \times 10^{-11}~\reciprocal{\second}$
			at $W = 1~\eV$,
		$1.5 \times 10^{-18}~\reciprocal{\second}$
			at $W = 1.7~\eV$,
		\etc
	The point is that
		solid state theory mandates such creep,
	but it's untreated
		in any branch of science or engineering.
	Many of these details were put together in
		mundane \texttt{arxiv.org} articles
		\cite{PrasadArxiv},
	to be eventually compiled into
		a comprehensive paper.
	Thanks to NASA/JPL's continued portrayal of the anomaly
		as acceleration,
	\emph{all} other explanations tested or offered
		have been more obvious or exotic,
			and totally fruitless
		(\cf \cite{Anderson2002}).
	}. 

\Subsection{ddm}{Distance division multiplexing\texttrademark}

\unc
Almost all the applications to be discussed
critically depend on
	the fundamental separation of sources
	enabled by
		the present wave effect.
Using the notation of quantum theory,
we may denote
	incoming signals by ``kets'' $\ket{}$,
and
	the receiver states
		as ``bras'' $\bra{}$,
and rewrite
	equation (\ref{e:ordorth}) as
	$
	\iprod{\widehat\omega}{\omega, r}
	=
		e^{ikr}
		\iprod{\widehat\omega}{\omega}
	$.
Equation (\ref{e:modorth})
then becomes
\begin{equation} \label{e:qshift}
\begin{split}
	\iprod{
		\widehat\omega,
		\frac{d \widehat\omega}{dt}
	}{\omega, r}
	&\equiv
		\bra{\widehat\omega}
		H
		\ket{\omega, r}
	=
		e^{ikr\Delta(r)}
		\,
		\iprod{
			\widehat\omega
		}
		{
			\frac{\omega}{
				\Delta(r)
				}
		}
\\
	&=
		e^{ikr\Delta(r)}
		\,
		\delta \left(
			\widehat\omega
			-
			\frac{\omega}{
				\Delta(r)
				}
			\right)
	\quad
	,
\end{split}
\end{equation}
where
	$\bra{\widehat\omega, d\widehat\omega/dt}$ and
	$\bra{\widehat\omega}$
are
	the modified and original states of the receiver,
	respectively.
The sampling clock and grating modifications
then correspond to the operator $H$
\begin{equation} \label{e:qop}
	H \ket{\omega, r}
	=
		e^{ikr\Delta(r)}
		\,
		\ket{\frac{\omega}{\Delta(r)}}
\end{equation}
of an incoming wave state
	$\ket{\omega,r}$.
These equations attribute the shift to
	the incoming value $\omega$,
instead of
	to $\widehat\omega$ selected instantaneously
since for the operator formalism,
	$\bra{\widehat\omega}$ must represent
		an eigenstate resulting from observation.

\begin{figure}[ht]
	\centering
	\psfig{file=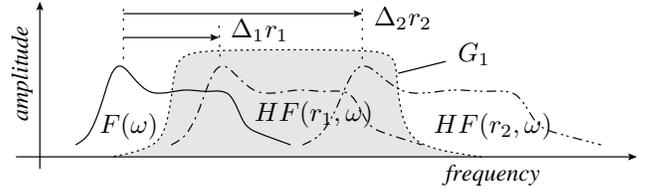}
	\caption{Separation by source distance}
	\label{f:DDM}
\end{figure}

\unc
Equations (\ref{e:zshift}) and (\ref{e:qop})
both say that as in the Doppler case,
	the shift is proportional
		at each individual frequency $\omega$.
This also means that
	the spectrum expands by the factor $\Delta$,
as illustrated in Fig.~\ref{f:DDM}
for the case of
	a common signal spectrum
		$F(\omega)$
emitted by two sources
	at distances $r_1$ and $r_2 > r_1$,
		respectively.
For a rate of change factor $\alpha$
	applied at the receiver,
the signals will shift and expand by
	$\Delta_1 \equiv (1 + \alpha r_1)$
and
	$\Delta_2 \equiv (1 + \alpha r_2) > \Delta_1$,
respectively.

\unc
Despite the increased spreading of the spectrum,
there is opportunity for
	isolating the signal of a desired source
if its shifted spectrum comes out
	substantially separated
	from its neighbours,
as shown.
We could, for instance, apply
	a band-pass filter $\widetilde{G}_1$
to the overall shifted spectrum 
	$H F_1 + H F_2$,
such that
	$
	\widetilde{G}_1 ( H F_1 + H F_2 )
	\approx \widetilde{G}_1 H F_1 .
	$
Writing $H$ as a function of $\alpha$,
	by equations (\ref{e:modorth}) and (\ref{e:qop}),
we get
\begin{equation} \label{e:negH}
	H^{-1}(\alpha) = - H(\alpha)
	\quad
	.
\end{equation}
We could apply a combination of
	mixing and frequency-modulation operations
to shift and compress
	$\widetilde{G}_1 H F_1$ back to $\approx F_1(\omega)$,
or just apply
	a second $H$ with a negative $\alpha$.
In general,
a set of distance-selecting
	``projection operators''
	$H^{-1} \widetilde{G}_i H$,
can thus be defined by
	the conditions
\begin{equation} \label{e:selection}
	H^{-1} \widetilde{G}_i H
		\,
		\sum_j F(r_j, \omega)
	\approx
		F(r_j, \omega)
\text{~or~}
	H^{-1} \widetilde{G}_i H
	\approx
		\delta_{ij}
\end{equation}
With base-band prefilters
	$G_i F_i \approx F_i$,
they yield
\begin{equation}
	H^{-1} \widetilde{G}_i H
	=
		\delta_{ij} G_j
\quad\text{so that}\quad
	\widetilde{G}_i H
	= 
		H G_i
	\quad
	.
\end{equation}

\unc
As $H$ is parametrized by $\alpha$
	independently of $i$,
	$
	H\widetilde{G}_i H
	$
provides spectral separation
	without prior knowledge of $r_i$.
It is also independent of signal content,
wherein
	the source coordinates could be supplied
		or included via modulation,
but since $H$ depends only on
	the path phase contribution,
it provides separation at
	a more basic level.
Coordinate-based spread spectrum coding
	could be combined,
for example,
to differentiate sources
	at the same distance,
as an alternative to or to improve over
	current phased array antennas.

\unc
It has been further suggested \cite{Prasad2005}
that DDM raises
	the theoretical capacity of a channel to
	$\sim 2 \mathcal{C} L / \lambda \gg \mathcal{C}$,
where
	$\mathcal{C}$ is the traditional Shannon capacity
		for a source-receiver pair using the channel,
by admitting
	additional sources into the channel
		at intermediate distances.
Further,
if combined with
	directional selectivity say from
		phased array antennae,
we would get true
	\emph{space-division multiplexing}
		[\ibid].

\Section{mil}{Military and intelligence applications} 

\unc
The above theory represents a long overdue
	rigorous analysis of
		the spectral decomposition of waveforms
	by a real, and therefore imperfect, instrument.
The nature of the error was inferred, as mentioned,
from an immense gamut of
	astronomical,
	planetary
and
	terrestrial data,
spanning all scales of range
	ever measured.
A natural occurrence of the effect
	subsequently deduced from creep theory
		(footnote, page \pageref{foot:cosmos})
could have been contradicted
	in at least one set of data,
but remains totally consistent%
	\footnote{
	\func
	Consistency with GPS datasets
		has also been recently verified.
	Calculations from GPS base stations data indicate
		an ongoing rising of land
	at a median rate of
		$
		1.6
		~\milli\metre\per\yyear
		$
		(see \texttt{http://ray.tomes.biz}),
	large enough to easily accommodate
		a nontectonic apparent expansion of the earth
	at $H_0 \times 6.371~\kilo\metre
		\approx 0.437~\milli\metre\per\yyear$,
	for the natural occurrence.
	}. 
Instead, we clearly face
	a converse problem in physics itself,
that its patterns
	have been so well explored on earth
that only astronomical tests can expose
	remaining shortcomings
	\cite{Anderson1998}.
The known laws of physics only reflect
	understood mechanisms,
like 
	interference and the Doppler effect
and only partially%
	\footnote{
	\func
	E.g.,
	diffractive corrections in
		astrophysics and quantum field theories
	are limited to
		the Fraunhofer and Fresnel approximations,
	both being limited, by definition, to
		total deflections of $\le \pi/2$. 
	Cumulative diffractive scattering would contribute,
		as known in microwave theory,
		a decay $e^{-\sigma r}$ to the propagation.
	This makes \emph{all wavefunctions}
		in the \emph{real universe} Klein-Gordon eigenfunctions,
	and gives light
		a rest-mass...
	}. 
We once again face a need to lead physics
	by engineering%
	\footnote{
	\func
	Recalling the classic case of thermodynamics.
	The present effect 
	implies that photons,
		traditionally viewed as indivisible and immutable,
			are reconstituted
		by every real, imperfect, receiver.
	Though it is yet to be demonstrated with light,
	the theory of Sections \ref{s:theory} and \ref{s:impl}
		is fundamental and does not permit
			a different result for quanta.
	}.\xspace

\Subsection{shift}{Precision, power-independent frequency shifting} 

\unc
This application would itself be
	a direct, visible test of the effect,
the idea being that
	the method could be applied also to
a proximal source
	at a precisely known distance
in order to uniformly scale its spectrum
	with great accuracy.
Accuracy is expected because
	the source distance
can be accurately set
	by simple mechanical means,
independently of
	the exponential control of $\alpha$.
The mechanism would be
	the first means for shifting frequencies
that is independent of
	the nature and energy of the waves,
		and their frequencies and bandwidth,
yet realizable by
	a mundane, static means.
The main difficulty is the magnitude of $\alpha$
	needed for source distances of under $1~\metre$,
but it should be
	eventually solvable.

\snf
Likely applications include
transformation of high power
	modulated $\giga\hertz$ carriers to
		$\tera\hertz$ or optical bands,
and	efficient, tunable
		visible light, UV or even X- or $\gamma$-rays,
all
	with controlled coherence
and
	without nonlinear media.

\ifthenelse{\boolean{extended}}{
\cbstart
\snf
Two forms of implementation are being explored,
	the variable grating scheme of
		Section \ref{ss:diffract},
	exploiting
		the magnetostrictive ``smart material'' Terfenol-D,
and
	the path variation method of
		Section \ref{ss:phaseshift}
	using a liquid crystal or a photorefractive medium
		for a faster modulation
	than seems possible with
		the Faraday effect in glasses.
Another aspect being studied is
	folding of the source path
using
	an optical fibre or a transmission line,
so as to ensure realizability of
	possibly several metres in a compact package
and
	at least $1~\metre$ in a single chip.
The Terfenol-D design,
	prepared for NSF SBIR/STTR Solicitation 05-557,
envisaged
	a grating etched directly on
		the side of a Terfenol-D medium,
and was expected to provide
	shift factors $z \ge 2$
at least with
	a strong, distant, commonly available source,
	\viz the sun,
as a first step of
	optical validation.
\cbend
}{}


\Subsection{radar}{Monostatic ranging and passive radar} 

\snf
This application was immediately envisaged
	when the effect was first actually suspected%
	\footnote{
	\func
	The idea of such an effect had been disclosed,
	by appointment, to an IP attorney
		on the morning of 2001.9.11.
	The mechanism itself (Section \ref{s:theory})
		was uncovered only much later in 2004.
	}, 
taking from the well known notion of
	the cosmological distance scale
		available from the Hubble redshifts.
By providing
	a ``virtual Hubble flow'' view
that can be activated and scaled
	at the receiver's discretion,
the present methods enable
	ranging, or distance measurement,
of any source
	that can be seen or received,
	at only half
		the round-trip delay,
and
	zero transmitted power.

\snf
To compare,
as traditional active radar depends on
	round trip times (RTT),
its range is fundamentally limited 
	by the transmitter power. 
While Venus and Mars have been explored by radar,
for example,
the ranging of other planets
	and of astronomical objects in general
is largely limited
	to spatial triangulation,
including
using the earth's orbit itself
	as the baseline.
Tracking of orbiting satellites 
places similar demands on
	the radar transmitter power $P$,
as the range $R$ is governed by
	the ``radar power law''
	$P \propto R^4$.
This makes
	current (active) radars generally bulky.
The power law is relaxed for
	cooperative targets with transponders,
		to $P \propto R^2$,
where $P$ denotes
	transponder power,
and the range is obtained from the RTT of
	a transponded signal%
	\footnote{
	\func
	This is generally what's used for deep space probes
		\cite{Bender1989,Vincent1990}.
	}. 
The present methods make this lower power law
	available for all targets,
and also eliminate the need for
	a spatial baseline for parallax,
as the baseline is now given by
	the range of variation of $\alpha$
		at the receiver.
As shown in Section \ref{ss:scale},
a very large baseline seems to be indeed possible
	with DSP alone
for both terrestrial
	and earth orbit distances.

\snf
Current passive radar technology,
	like Lockheed's Silent Sentry,
involves
	coherent processing of the received scatter
of energy from
	television broadcast
	(see US Patent 3,812,493, 1974)
and
	cellular base stations.
The accuracy is again dependent on
	information representing overall trip times,
and hence entails
	coherent processing.

\snf
Accurate ranging is now possible via
	triangulation using temporal parallax,
as explained
	in Section \ref{ss:diff}.
The difference is that
	coherent processing can provide
		imaging
	with subwavelength precision,
but
	triangulation would be simpler
	for ranging and tracking.
We still need to illuminate silent targets,
but the processing could be simplified
	for the same accuracy.
The accuracy of tracking could improve
	for radiating targets
given
	the lower overall trip time.
We can also use
	temporal triangulation
to simplify traditional radar
	and as a fast, efficient cross-check.

\Subsection{jam}{Overcoming interference, jamming and noise} 

\snf
The capability for separating sources
	regardless of the signal content or modulation
is also a guarantee that
	a desired signal can be isolated
		from any interfering signals
at the same frequencies and
	bearing the same modulations or spread-spectrum codes,
so long as
	the interfering sources are
		at other directions or distances
	from the receiver.

\snf
Simulation shows that
	even signals in destructive interference
		can be separated.
The criterion for separation is simply
	the bandwidth to distance ratio:
denoting the low and high frequency bounds of
the signal bandwidth $W$ by
	$\mathcal{L}$ and $\mathcal{H}$, respectively,
each of equations (\ref{e:zshift}) or (\ref{e:modorth})
	implies
\begin{equation}
	(1 + \alpha r_i) \mathcal{H}
		\le
	(1 + \alpha r_{i+1}) \mathcal{L}
\end{equation}
as the condition for separation between
	the $i$th and $(i+1)$th sources,
		assuming $r_i \le r_{i+1}$
(see Fig. \ref{f:DDM}).
This means
\begin{equation}
	\alpha r_i
	\ge
		\frac{W}{(\delta r_i/r_i) \mathcal{L} - W}
	=
	\left[
		\frac{\delta r_i}{r_i}
		\frac{\mathcal{L}}{W}
		-
		1
	\right]
\end{equation}
where $W = \mathcal{H} - \mathcal{L}$
	and $\delta r_i = r_{i+1} - r_i$.
As separation is only possible for
	$\alpha > 0$,
we need to have
	$\delta r_i > r_i W / \mathcal{L}$.
It then follows, however,
that sufficiently narrow subbands of
	the total received signal
	would be ``source-separable''
even when the condition should fail
	for $W$ as a whole.

\snf
All of the limitations are thus purely technological,
\eg
	how many and how narrow subbands
		we can construct,
	how linear we can make
		the subband and band-pass ($\widehat{G}$) filters
	(since phase envelope distortions will alter the shift),
and
	how large the stop-band rejection
		we can get.
The stop-band rejection ratio is important
	in overcoming jamming,
and the separation as such should suffice, in principle,
	for filtering out truly extraneous noise
		like lightnings.

\Subsection{isar}{Incoherent aperture synthesis} 

\begin{figure}[ht]
	\centering
	\psfig{file=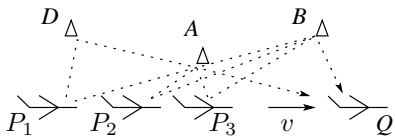}
	\caption{Aperture synthesis}
	\label{f:sar}
\end{figure}

\unc
In a conventional SAR,
multiple targets or features are imaged
	from the echos received on a moving platform,
		typically an aircraft or satellite,
	as depicted in Fig.~\ref{f:sar}.

\unc
If the onboard oscillator could maintain coherence for
	the duration of the flight,
\ie not suffer random phase shifts,
an entire target region could be imaged
	by a simple Fourier transform,
after correcting for
	the Doppler shifts in the received echos
		due to the radar's own motion.
At each point $Q$
	on the flight path,
echos are received from multiple features
	$A$, $B$, \etc
and corresponding to 
	multiple previous locations
		$P_1$, $P_2$, ...
	of the transmitter.
Fourier techniques and
	their optical holographic counterparts
are most efficient
	for unravelling the information.

\snf
As stated in Section \ref{ss:radar},
the fundamental reason for coherent processing
	is our current dependence on RTT of radar echos
		for the range information.
The dependence is obviated by
	the wave effect.
In the example scenario,
echos received at $Q$
	from target features differing in direction,
	like $A$ and $B$,
can be separated by the phase information 
	from an antenna array.
The echos from the same direction,
	such as from $A$ and $D$,
can now be similarly separated
	using the methods of
		Section \ref{s:impl}
--
in particular,
the changing delay transform
	of equation (\ref{e:kdft})
can be applied to 
	recordings of the received echo waveform.

\snf
Coherent processing ordinarily has
	two advantages
that would seem to be impossible
	in any other approach:
it tends to curtail noise and is generally precise
	to within a wavelength.
We would get suppression of external noise,
	as discussed in Section \ref{ss:jam},
and subwavelength resolution,
	as shown in Section \ref{ss:diff},
so that applications including
	hand-held short-range carrier-deck operation,
for which coherent processing is difficult,
	become feasible.

\ifthenelse{\boolean{extended}}{
\Section{sim}{Simulation and testing} 

\cbstart
\unc
The DSP approach of Section \ref{ss:dsp}
	was first tried in 2001 using Octave
	(an open source equivalent of Matlab),
on several \texttt{.wav} files
	that came with Windows and Linux distributions.
The expectation of discernible shifts
	seems to hold best for
the simple waveform of \emph{pop.wav}
	(Fig.~\ref{f:PopWave}),
with a fundamental frequency
	below $100~\hertz$.
Fig.~\ref{f:PopShift} shows
	the corresponding spectra:
line 1 (in red) is
	the spectrum of the unmodified samples,
and lines 2 (in green) and 3 (in blue) are
spectra obtained by
	interpolating the samples to simulate
		time-varying sampling rates
	according to Fig.~\ref{f:TimeDomain}.
The shifted spectra are spread out
	but otherwise preserved,
as expected,
	at the lower frequency end.
The harmonic peaks
	near $310~\hertz$ and $420~\hertz$
		are lost,
and
	the shifted spectra are increasingly weaker
--
both behaviours are expected as
	the interpolation reduces the number of samples.
\cbend

\begin{figure}[htb]
	\centering
	\psfig{file=popwaveform.eps, height=1.6in}
	\caption{Sample acoustic waveform}
	\label{f:PopWave}
\end{figure}

\begin{figure}[htb]
	\centering
	\psfig{file=pop.eps, height=1.6in}
	\caption{Original and shifted spectra}
	\label{f:PopShift}
\end{figure}

\cbstart
\unc
With other audio samples,
and with a newer simulator developed in Java
	especially to study
the performance of subband filtering 
	described in Section \ref{ss:jam}
		for antijamming,
the frequency scaling of acoustic spectra
	is found to be less distinct.
Rounding and aliasing problems are exacerbated
	by the nonlinear sample interpolation,
but the main cause seems to be
	the distance to the microphone
		in the recording process
being
	equal to or smaller than
		the distribution of the recorded sources,
so that
	$\Delta(\omega t)$ dominates over
		the path length contribution
	$\mathbf{k} \cdot \Delta\mathbf{r}$
	(equation \ref{e:dphase}).
This can be solved by introducing
	a fixed delay corresponding to a large source distance,
and the expected effect of increased shifts
	was verified
		in the Octave tests.
In the new simulator,
different subbands in most audio \texttt{.wav} files
	are found to scale by different extents,
consistent with
	the interpretation of source spread.
Nevertheless,
tests with actual, well-sampled sonar data
	are needed to validate and characterize
		the effect adequately
	for use with sound.
The source spread problem would be worse for
	longer radio waves,
making
	the spectral phase gradients too small
		to be usable
	at long wavelengths.
This problem should vanish at short wavelengths,
and in particular,
	at optical wavelengths
where individual photons could be traced, in principle,
	to atoms that emitted them.
The source spread would likely arise again in
	laser output,
in proportion to the spread of time spent by the photons
	within the lasing cavity,
and this too
	remains to be tested.
\cbend

\cbstart
\unc
The new simulator continues to evolve
	as the lessons learnt are incorporated
		into the code.
Development is slow because
	the research is as yet unfunded
and being done in spare time spread across
	complementary research efforts
		that led to this discovery.
There were programming errors causing noise
	and limiting the range of $\alpha$
		in the simulation,
for instance,
and were identified and corrected only after
	the initial presentation and demonstration
		at WCNC 2005.
The simulator still uses sample interpolation
	(Section \ref{ss:dsp}),
which needs to be replaced with
	the more robust path variation scheme
		of Section \ref{ss:phaseshift}.
\cbend

\cbstart
\unc
Direct validation with real RF data and live feeds,
	for the DSP methods,
and with
	a Terfenol-D or liquid crystal device
		for optical wavelengths,
remain to be done,
the main limitation in each case being
	the lack of current skill with
		electronic circuit design,
	and the time,
	to implement them.
Nevertheless,
there is an enormous gamut of
	astronomical and geological evidence
consistent with
	the effect at optical wavelengths
and the uncorrected natural cause mentioned,
as reviewed in
	the WCNC presentation,
and the capability to recover
	signals even from total destructive interference,
also demonstrated at WCNC,
	should be hopefully sufficient to motivate
		independent validation efforts
and funding
	for full scale development and exploitation
		of this new physics.
The WCNC presentation and
	a demonstration of separation of an FM signal
		from similar interfering signals
	using the new simulator
have accordingly been made available online
	at \texttt{http://www.inspiredresearch.com} .
\cbend
}{}

\section*{Acknowledgement}

\unc
I thank
Asoke K. Bhattacharyya,
	for introducing me to
		inverse scattering in pulsed radar
	in 1984,
and thus to
	a key insight leading to this discovery.

\begin{raggedright}

\end{raggedright}

\begin{thebibliography}{1}

\bibitem{Prasad2005}
V~Guruprasad.
\newblock {R}elaxed bandwidth sharing with {S}pace {D}ivision {M}ultiplexing.
\newblock In {\em Proceedings of the IEEE Wireless Communication and Networking
  Conference}, March 2005.

\bibitem{Parker1969}
L~Parker.
\newblock Quantized fields and particle creation in expanding universe - {I}.
\newblock {\em Phys Rev}, 183(5):1057--1068, 25~Jul 1969.

\bibitem{Anderson2002}
J~D Anderson et~al.
\newblock {S}tudy of the anomalous acceleration of {P}ioneer 10 and 11.
\newblock {\em Phys Rev D}, 65, Apr 2002.
\newblock Also: report LA-UR-00-5654 and gr-qc/0104064, Apr 2001.

\bibitem{PrasadArxiv}
V~Guruprasad (unpublished).
\newblock astro-ph/9907363 (analysis of the {P}ioneer 10/11 anomalies, 1999),
  gr-qc/0005014 (``prediction'' of the cosmological acceleration , 2000).
\newblock At http://www.arxiv.org, both cited in \cite{Anderson2002}.

\bibitem{Anderson1998}
J~D Anderson et~al.
\newblock Indication from {P}ioneer 10/11, {G}alileo and {U}lysses data of an
  apparent anomalous, weak, long-range acceleration.
\newblock {\em Phys Rev Lett}, Oct 1998.

\bibitem{Feynman}
R~P Feynman, R~Leighton, and M~Sands.
\newblock {\em The {F}eynman {L}ectures on {P}hysics}.
\newblock Addison-Wesley, 1964.

\bibitem{Bender1989}
P~L Bender and M~A Vincent.
\newblock Small {M}ercury {R}elativity {O}rbiter.
\newblock Technical Report N90-19940 12-90, NASA, Aug 1989.

\bibitem{Vincent1990}
M~A Vincent and P~L Bender.
\newblock Orbit determination and gravitational field accuracy for a {M}ercury
  transponder satellite.
\newblock {\em J Geophy Res}, 95:21357--21361, Dec 1990.

\end{thebibliography}
\end{document}